\documentclass[a4paper,11pt]{article}
\usepackage{pos}
\usepackage{bm}
\usepackage{natbib}
\usepackage{verbatim}
\title{Signal Model and Energy Reconstruction for the Radio Detection of Inclined Air Showers in the 50-200 MHz Frequency Band}
\ShortTitle{Radio Signal Model and Energy Reconstruction for Inclined Air Showers in 50-200 MHz}

\author[a]{Lukas Gülzow}
\author[a,b]{Ralph Engel}
\author*[a,c]{Tim Huege}
\author[a]{Markus Roth}
\author[d]{Felix Schlüter}

\affiliation[a]{Institute for Astroparticle Physics (IAP), Karlsruhe Institute of Technology, Karlsruhe, Germany}
\affiliation[b]{Institute of Experimental Particle Physics (ETP), Karlsruhe Institute of Technology, Karlsruhe, Germany}
\affiliation[c]{Astrophysical Institute, Vrije Universiteit Brussel, Brussels, Belgium}
\affiliation[d]{Université Libre de Bruxelles, Science Faculty CP230, B-1050 Brussels, Belgium}

\emailAdd{lukas.guelzow@kit.edu}

\abstract{The radio emission of cosmic-ray air-showers changes significantly depending on parameters like signal frequency, magnetic field configuration and observing altitude.
We use CoREAS simulations to adapt an existing signal model for the radio emission of inclined showers in the 30-80 MHz frequency band to the wide $50-200\,\mathrm{MHz}$ band.
Our model uses a parametrisation of the charge excess fraction to isolate the geomagnetic emission component.
We reconstruct the geomagnetic radiation energy by fitting a lateral distribution function, provided by the model, to the geomagnetic energy fluence distribution of the shower.
After we correct for the shower geometry and air density, we correlate the radiation energy with the electromagnetic energy of the shower.

We show that the method intrinsic energy resolutions $<5\%$ for the sites of the Pierre Auger Observatory and GRANDProto300.
For GRANDProto300, we test the reconstruction with simulations of a realistic, sparse antenna grid and with added noise, and find an energy resolution of $<10\%$ with negligible bias.
We do a similar study for a much larger array of $10,000\,\mathrm{km^2}$ with $1\,\mathrm{km}$ antenna spacing.
We find an intrinsic energy resolution of $<10\%$.}

\ConferenceLogo{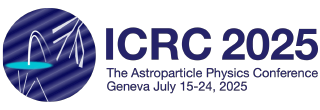}

\FullConference{%
  39th International Cosmic Ray Conference (ICRC 2025)\\
  July 15-24 2025\\
  CICG - International Conference Centre - Geneva, Switzerland
}

\begin{document}
\maketitle

\section{Introduction}
\label{sec:intro}

The signal model and energy reconstruction of inclined air showers developed for the AugerPrime Radio detector in the frequency band of $30-80\,\mathrm{MHz}$~\cite{Model-paper} has been used with great success on both simulations and data~\cite{Pont_ICRC25}.
On simulations with a sparse antenna grid, it achieves an intrinsic $<5\%$ energy resolution with no significant biases.
For fully realistic simulations, exclusively using information available in real measurements, the resolution only decreases to $\sim 6\%$ while remaining free of biases~\cite{Huege-performance}.
This warrants an adaptation of this method to other frequency bands and for different experiments and sites.
In this work, we adapt this method to the $50-200\,\mathrm{MHz}$ frequency band in which the Giant Radio Array for Neutrino Detection (GRAND) operates~\cite{GRAND:2018iaj}.

The radio emission of air showers is composed of a geomagnetic and a charge excess component~\cite{Huege-review}.
For inclined air showers, the emission pattern exhibits geometry-dependent asymmetries.
We deal with these asymmetries with a so-called early-late correction~\cite{Huege:2018kvt}.
We describe the fraction of the charge excess $a_\text{ce}$ of the total emission with a parametrisation and use it to isolate the geomagnetic component of the energy fluence $f_\text{geo}$.
We fit a 1-dimensional geomagnetic lateral distribution function (LDF) to $f_\text{geo}$.
By integrating the resulting function, we calculate the geomagnetic radiation energy $E_\text{geo}$ of the shower.
After applying corrections for the angle to the magnetic field and for air density effects, this allows for a reconstruction of the electromagnetic energy $E_\text{em}$.
To enable the energy reconstruction for sparse antenna arrays, we parametrise the shape parameters of the LDF with respect to the distance $d_\text{max}$ from the shower core to the shower maximum.

In this work, we tune the model for two sites using two simulation libraries generated with CoREAS~\cite{Huege:2013vt}.
We choose the site of the Pierre Auger Observatory in Argentina, where both the GRAND prototype detector GRAND@Auger~\cite{G@A_ICRC25} and an upcoming array building on SKALA antennas~\cite{SKALA_ICRC25} currently operate in higher frequency bands.
The second site is the site of the GRAND prototype detector GRANDProto300 (GP300) near Dunhuang in China~\cite{GP300_PoS}. 
The library for Argentina is the same as the one used in reference~\cite{Model-paper}, filtered to the $50-200\,\mathrm{MHz}$ frequency band.
The antenna grids in both simulation libraries are star-shaped and centred on the Monte-Carlo shower core for an idealised characterisation of the highly inclined air showers.
For Argentina, the "Malargue October" atmosphere is used and, for China, we use the "Dunhuang" model.
The atmosphere models are provided by the \texttt{radiotools} package~\cite{Glaser_2019}).

The higher and wider frequency band brings with it differences in the radio emission.
The so-called Cherenkov ring of the radio emission pattern is more pronounced and the signal coherence decreases due to the shorter wavelengths.
In addition, the much stronger magnetic field in China exacerbates this loss of coherence~\cite{Chiche:2024yos}.
We address these effects and adapt the reconstruction accordingly in this work.
With Argentina and China serving as extremes of the magnetic field strength on Earth, we aim to show that the model is adaptable for any magnetic field configuration.

We use two CoREAS simulation libraries for the GP300 site with realistic, sparse antenna layouts and with added artificial noise to benchmark the performance of the energy reconstruction.
The three simulation libraries for the GP300 site have been created specifically for this work.

In Section~\ref{sec:model}, we cover the steps of the signal model.
In Section~\ref{sec:recon}, we describe the energy reconstruction and the compensation for the loss of coherence.
We display the reconstruction performance for Argentina and China on multiple sets of simulations with different antenna array configurations in Section~\ref{sec:performance}.
Finally, we draw our conclusions in Section~\ref{sec:conclusions}.

\section{Signal Model}
\label{sec:model}

%
For inclined air-showers, the radio emission pattern exhibits geometry-dependent asymmetries.
Our model corrects for these asymmetries with a so-called early-late correction~\cite{Huege:2018kvt}.
From there, we calculate the geomagnetic energy fluence $f_\text{geo}$ from the component of the electric field polarised in the $\Vec{v}\times\Vec{B}$ direction and a parametrisation of the charge excess fraction of the total emission.
The charge excess fraction in this case is defined in terms of the energy fluence.

\begin{figure}
\begin{center}
	\includegraphics[clip, trim=13.5cm 0.5cm 0.5cm 3.5cm, width=0.8\columnwidth]{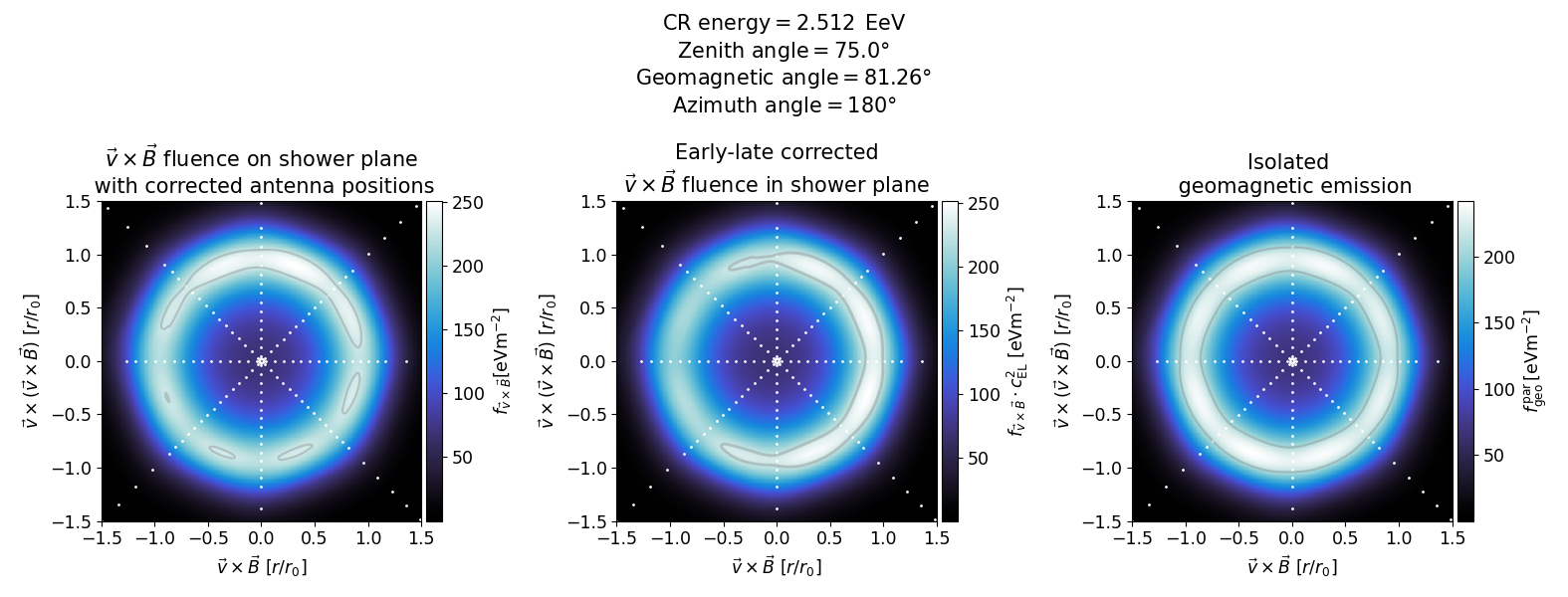}
    \caption{The first two steps of our signal model which eliminate geometric asymmetries and to isolate the geomagnetic energy fluence, respectively. Shown for an example simulation of the GRAND@Auger site with a star-shaped antenna pattern (white dots). The regions with $90\%$ or more of the maximum energy fluence are outlined in grey. Axis distances $r$ are normalised with respect to the Cherenkov radius $r_0$. \textbf{Left:} Early-late correction factors $c_\text{EL}$ applied to energy fluence and antenna positions. \textbf{Right:} Parametric geomagnetic fluence $f_\text{geo}^\text{par}$.}
    \label{fig:footprints}     
\end{center}
\end{figure}

By applying these steps, we find the symmetric fluence pattern of the geomagnetic emission.
The geomagnetic fluence $f_\text{geo}^\text{par}$ and the parametrisation of the charge excess fraction $a_\text{ce}$ are given by Eqs. (4.10) and (4.12) in reference \cite{Model-paper}.
We refit the charge excess parametrisations
\begin{equation}
a_\text{ce}^\text{ARG}=\left[0.302-\frac{d_\text{max}}{729\,\mathrm{km}}\right]\cdot\frac{r_\text{axis}}{d_\text{max}}\cdot\text{exp}\left(\frac{r_\text{axis}}{682\,\mathrm{m}}\right)\cdot\left[\left(\frac{\rho_\text{max}}{0.422\,\mathrm{kg}\,\mathrm{m}^{-3}}\right)^{2.98}+0.178\right],
\label{eq:a_ce_param_auger}
\end{equation}
\begin{equation}
a_\text{ce}^\text{CHN}=\left[0.229-\frac{d_\text{max}}{1106\,\mathrm{km}}\right]\cdot\frac{r_\text{axis}}{d_\text{max}}\cdot\text{exp}\left(\frac{r_\text{axis}}{614\,\mathrm{m}}\right)\cdot\left[\left(\frac{\rho_\text{max}}{0.668\,\mathrm{kg}\,\mathrm{m}^{-3}}\right)^{1.43}+0.166\right],
\label{eq:a_ce_param_china}
\end{equation}
for the higher-frequency model for Argentina (ARG) and China (CHN), respectively.
In Fig.~\ref{fig:footprints}, we show the application of the signal model up to this point for the $50-200\,\mathrm{MHz}$ frequency band on the fluence footprint in the shower plane.
We show the early-late corrected $\vec{v}\times\vec{B}$ fluence $f_{\vec{v}\times\vec{B}}$ in the left panel and the now symmetric parametrised geomagnetic fluence $f_\text{geo}^\text{par}$ in the right panel.

In the next step, we fit a lateral distribution function (LDF) to the lateral geomagnetic fluence distribution.
We change the functional form of the LDF $f_\text{LDF}(r)$ to
\begin{equation}
f_\text{LDF}(r)=
    \begin{cases}
        \frac{E_\text{geo}}{E_0} \left[\text{exp}\left(-\left(\frac{|r-r_0|}{\sigma}\right)^{p_\text{inner}}\right)+\frac{a_\text{rel}}{1+\,\text{exp}\left(s\cdot\left(\frac{r}{r_0}-r_{02}\right)\right)}\right] & \text{for} \quad r<r_0,\\[15pt]
        \frac{E_\text{geo}}{E_0} \left[\text{exp}\left(-\left(\frac{r-r_0}{\sigma}\right)^{p(r)}\right)+\frac{a_\text{rel}}{1+\,\text{exp}\left(s\cdot\left(\frac{r}{r_0}-r_{02}\right)\right)}\right] & \text{for} \quad r\geq r_0.
    \end{cases} 
\label{eq:new_LDF}
\end{equation}
from the previous iteration of the signal model.
The shape of the LDF is characterised by a Gaussian and a sigmoid term.
It has seven shape parameters: the Gaussian peak position $r_0$, the Gaussian width $\sigma$, the amplitude of the sigmoid relative to the Gaussian $a_\text{rel}$, the sigmoid slope $s$, the sigmoid length scale $r_{02}$, and, lastly, the Gaussian exponent $p(r)$
\begin{equation}
    p(r) = 
    \begin{cases} 
        p_\text{inner} & r < r_0 \\
        2 \cdot (r_0 / r)^{b/1000} & r \geq r_0.
    \end{cases}
    \label{eq:gauss_slope}
\end{equation}
We fit $p(r)$ directly as $p_\text{inner}$ for $r<r_0$ and let it decrease slightly from 2 with an indirect fit of $b$ for $r\geq r_0$.
In the previous iteration, $p(r)$ is a constant for $r<r_0$.
We also directly determine the geomagnetic radiation energy $E_\text{geo}$ in the LDF fit with its normalisation $E_0$.
The position of the symmetry centre of the radio-emission footprint ("radio core") is a free parameter as well.
The radio core position is allowed to vary to accommodate its shift from the particle core of the shower due to refractive displacement~\cite{Schluter-refr_displacement}.
Its variation influences the value of $r$ for every antenna.

\begin{figure}
\begin{center}
	\includegraphics[clip, trim=2cm 0cm 0cm 1.34cm, width=0.8\columnwidth]{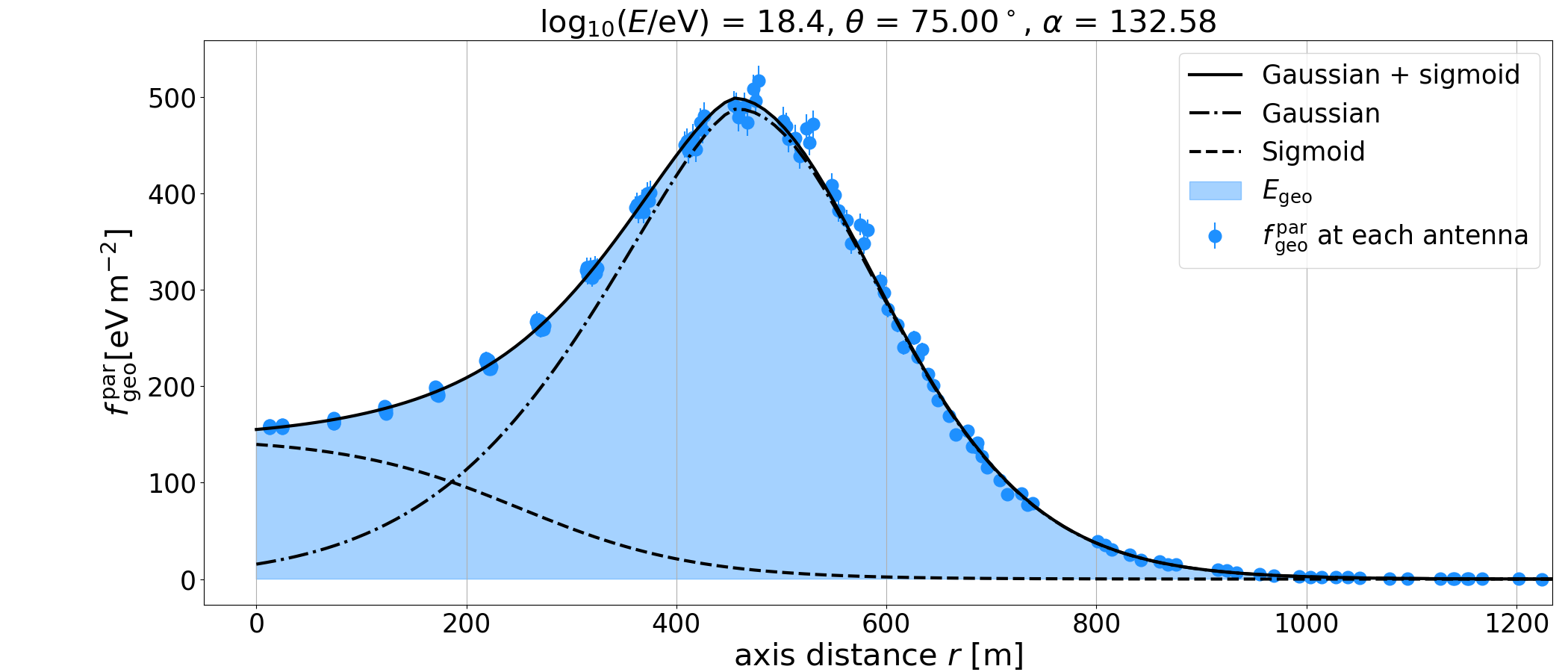}
    \caption{Fit of $f_\text{LDF}(r)$ (solid black line) to the geomagnetic fluence $f_\text{geo}^\text{par}$ (blue points) of an example air-shower simulation with a star-shaped antenna pattern using the properties of the GP300 site in China.
    We display the Gaussian and sigmoid components of the LDF as dot-dashed and dashed lines, respectively.
    The coloured area under the curve represents the geomagnetic radiation energy $E_\text{geo}$ of the air-shower.}
    \label{fig:integral}
\end{center}
\end{figure}

Fitting the LDF to the emission of an idealised air shower simulation works remarkably well (see Fig.~\ref{fig:integral}). 
However, it has too many free parameters to be stable for reconstruction using measurements with a sparse array. 
In many cases, especially for small zenith angles or low cosmic-ray primary energies, very few antennas have a sufficient signal-to-noise ratio.
To solve this problem, we parametrise all LDF shape parameters with the distanceto the shower maximum $d_{\max}$ .
This reduces the number of degrees of freedom to four: 
$E_\text{geo}$, $d_{\max}$ and the two radio core coordinates.
As such, the LDF fit only requires five antennas with signal to characterise the emission.

We parametrise the LDF iteratively.
We pick a shape parameter and observe its behaviour with respect to $d_{\max}$. 
Then, we fit a parametrisation to the shape parameter that characterises this behaviour. 
In the next iteration of the fit, we fix this shape parameter to its $d_{\max}$ parametrisation and pick another shape parameter to parametrise.
We repeat these steps until the LDF only depends on $E_\text{geo}$, $d_{\max}$ and the two core coordinates.
Differently from the other shape parameters, we fix the shape parameter $s$, describing the slope of the sigmoid component, to a constant.
This is the first step to ensure that the sigmoid is only significant inside the Cherenkov radius.

The shape parameters
\begin{equation}
    \sigma = \left(c_1 \cdot \left( \frac{d_{{\max}} - 5 \text{ km}}{\text{m}} \right)^{c_2} + c_3\right) \text{m},
    \label{eq:sigma_param}
\end{equation}
\begin{equation}
r_0=r_0^\text{pred} \cdot c_\text{par}(d_\text{max}) = 
r_0^\text{pred}\cdot \left(c_1 + \frac{d_{\max}}{c_2} + \frac{c_3}{d_{\max}^2}\right),
\label{eq:r0_param}
\end{equation}
represent, respectively, the width and radius of the Cherenkov ring.
$r_0^\text{pred}$ is the theoretical prediction for the radius of the Cherenkov ring using the zenith angle, $d_\text{max}$, observation height above sea level and the refractive index at sea level~\cite{Glaser_2019}.
We parametrise the remaining shape parameters directly with the $c_\text{par}(d_\text{max})$ term from Eq.~(\ref{eq:r0_param}) without a leading coefficient.
We display values for each parametrisation, and the order in which we fix them in Table~\ref{tab:params} for both Argentina and China.
\begin{table}[b!]
    \centering
    \begin{tabular}{l|ccc|ccc|}
         & \multicolumn{3}{c|}{Argentina} & \multicolumn{3}{c}{China} \\
        \hline
        Parameter & $c_1$ & $c_2$ & $c_3$ & $c_1$ & $c_2$ & $c_3$\\
        \hline
        $r_{02}$ & 0.666 & $771.9\,\mathrm{km}$ & $98.65\,\mathrm{km^2}$ & 0.586 & 1,176$\,\mathrm{km}$ & $166.3\,\mathrm{km^2}$\\
        $p_\text{inner}$ & 1.464 & 18,982$\,\mathrm{km}$ & $32.94\,\mathrm{km^2}$ & 1.541 & 3,810$\,\mathrm{km}$ & $64.99\,\mathrm{km^2}$\\
        $a_\text{rel}$ & 0.233 & 4,848$\,\mathrm{km}$ & $3.79\,\mathrm{km^2}$ & 0.281 & 2,173$\,\mathrm{km}$ & $34.78\,\mathrm{km^2}$\\
        $b$ & 282.2 & $2.73\,\mathrm{km}$ & 6,457$\,\mathrm{km^2}$ & 249.7 & $4.21\,\mathrm{km}$ & 14,561$\,\mathrm{km^2}$\\
        $\sigma$ & 0.027 & 0.805 & 61.97 & 0.035 & 0.770 & 62.94\\
        $r_0$ & 0.941 & 4,536$\,\mathrm{km}$ & $15.96\,\mathrm{km^2}$ & 0.818 & 1,776$\,\mathrm{km}$ & $36.82\,\mathrm{km^2}$\\
        \hline
    \end{tabular}
    \caption{Best fit values for the shape parameters of $f_\text{LDF}(r)$ for Argentina and China.
    We fix the parameters iteratively in the same order for both sites, indicated by the ordering of the table.
    We use the $c_\text{par}(d_\text{max})$ term of Eq.~(\ref{eq:r0_param}) for the first four parameters.
    $\sigma$ and $r_0$ are fixed to Eqs.~(\ref{eq:sigma_param}) and (\ref{eq:r0_param}), respectively.}
    \label{tab:params}
\end{table}

\section{Energy Reconstruction}
\label{sec:recon}

For the $50-200\,\mathrm{MHz}$ frequency band, we expect a loss of coherence from the radio emission due to the shorter wavelengths of the signal.
In addition, the magnetic field in China is almost 3x stronger than in Argentina, which lies within the South Atlantic Anomaly.
For China, this is the leading contribution to the loss of coherence since the electrons and positrons of the cascade experience a stronger Lorentz force and more easily exceed the coherence length of the emission.
Especially for the highest zenith angles, this leads to weaker signals reaching the antennas.
The loss of coherence also influences the polarisation signatures of the emission~\cite{Chiche:2024yos}.

\begin{figure}
	\includegraphics[clip, trim=1cm 1cm 2cm 0.75cm, width=0.5\columnwidth]{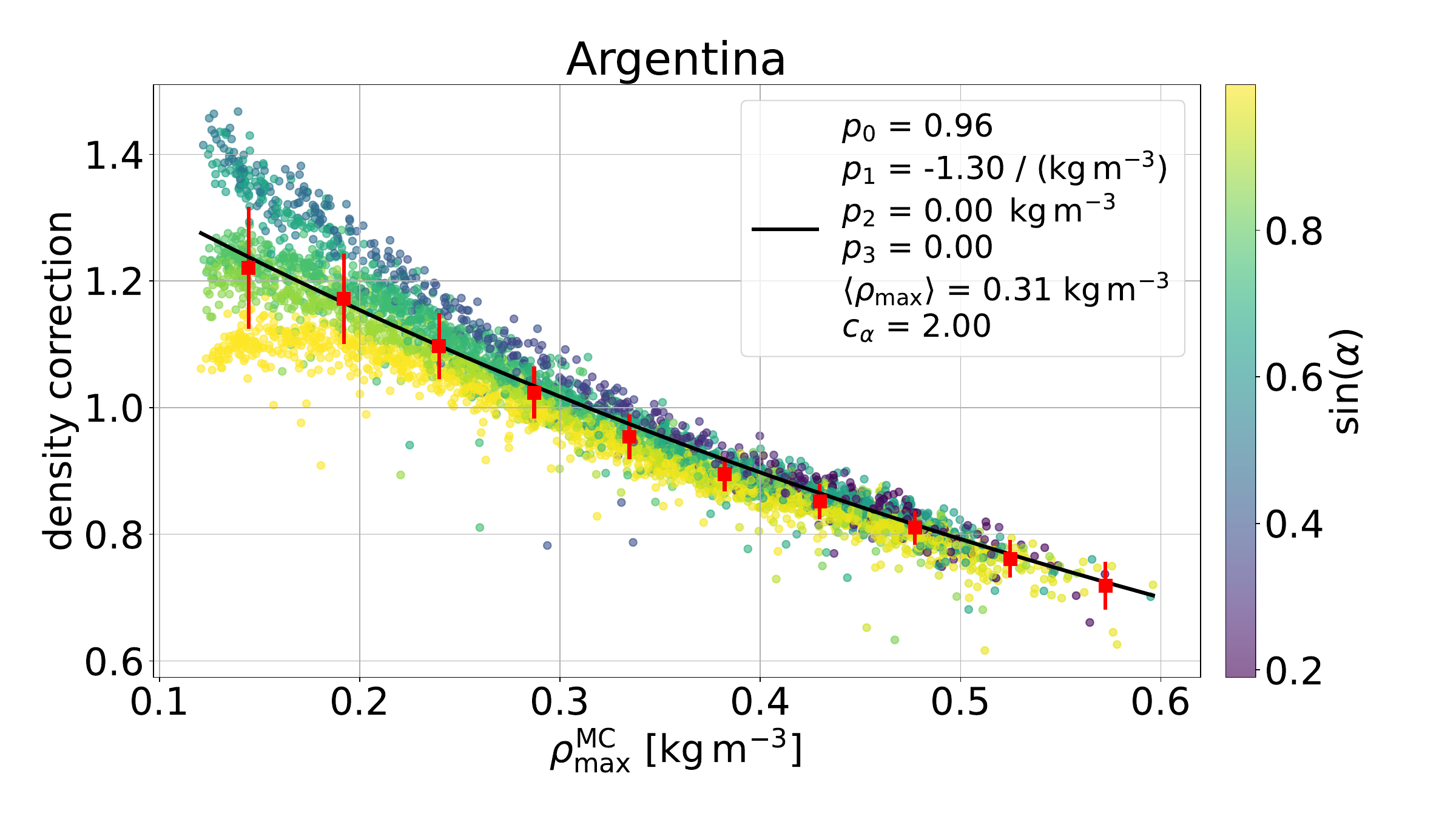}
    \includegraphics[clip, trim=1cm 1cm 2.4cm 0.75cm, width=0.5\columnwidth]{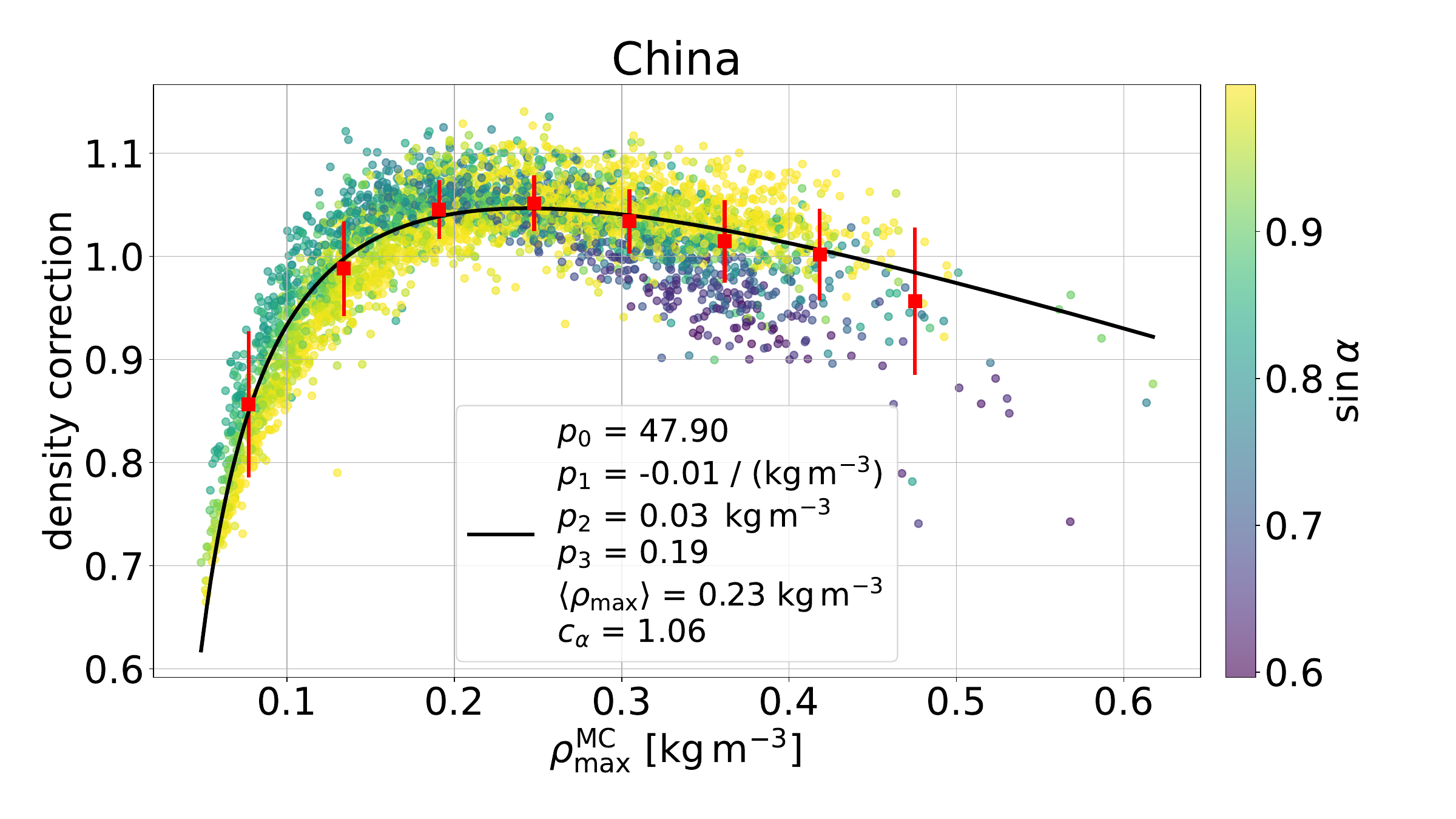}
    \caption{Modelled density correction term of Eq.~(\ref{eq:density_corr}) for the calculation of $S_\text{geo}$ plotted against $\rho_\text{max}$.
    The colour map shows the dependency on the geomagnetic angle $\alpha$.
    The black line is determined in a joint fit of Eqs.~(\ref{eq:density_corr}) and (\ref{eq:em}), which also provides values for $S_{19}$ and $\gamma$ (see Fig.~\ref{fig:intrinsic}).
    Red points and error bars show
    means and standard deviations, respectively.
    \textbf{Left:} Argentina. \textbf{Right:} China.}
    \label{fig:density_corr}
\end{figure}

In our model, the loss of coherence influences the density correction we apply to the radiation energy $E_\text{geo}$ to calculate the corrected radiation energy $S_\text{geo}$, which is independent of the shower geometry.
A signal with reduced coherence leads to a changed dependence on the geomagnetic angle $\alpha$, and causes $E_\text{geo}$ to fall off at low densities at the shower maximum $\rho_\text{max}$, e.g. high zenith angles where the cascade starts at high altitude.
To account for this, we modify Eq. (5.2) from reference \cite{Model-paper} with new parameters $c_\alpha$, $p_2$ and $p_3$.
This yields the following normalisation:
\begin{equation}
S_\text{geo}=\frac{E_\text{geo}}{\text{sin}^{c_\alpha}(\alpha)}\cdot\frac{1}{\left(1-p_0+p_0\cdot\text{exp}\left(p_1\cdot\left[\rho_\text{max}-\langle\rho_\text{max}\rangle\right]\right) - p_2\cdot \rho_\text{max}^{-1} + p_3\right)^2}.
\label{eq:density_corr}
\end{equation}
From $S_\text{geo}$, we can reconstruct the electromagnetic shower energy $E_\text{em}$
\begin{equation}
E_\text{em}=10\,\mathrm{EeV}\left(\frac{S_\text{geo}}{S_{19}}\right)^\frac{1}{\gamma},
\label{eq:em}
\end{equation}
with a power law with index $1/\gamma$. $S_{19}$ represents a typical geomagnetic radiation energy of an air shower with cosmic-ray primary energy of $10\,\mathrm{EeV}$ and an air density of $\rho_{\max}=\langle\rho_\text{max}\rangle$ at its shower maximum.
$\langle\rho_\text{max}\rangle$ is the average air density at the shower maximum of all simulations used to parametrise the LDF for the specific site. 
We determine the parameter values of $p_0$, $p_1$, $p_2$, $p_3$, $c_\alpha$, $S_{19}$ and $\gamma$ in a joint fit of Eqs.~(\ref{eq:density_corr}) and (\ref{eq:em}).
We show the best fit values for both sites as well as the respective values for $\langle\rho_\text{max}\rangle$ in the legends of the corresponding panels of Fig.~\ref{fig:density_corr}~and~\ref{fig:intrinsic}.

Fig.~\ref{fig:density_corr} shows the behaviour of the density correction term of Eq.~(\ref{eq:density_corr}) for the two sites.
For Argentina, we observe slight effects of coherence loss only for large values of $\text{sin}(\alpha)$ at low $\rho_\text{max}$ (i.e.~for large zenith angles).
As such, the new parameters do not have an influence.
For the stronger magnetic field in China, the loss of coherence is much more significant.

\section{Reconstruction performance}
\label{sec:performance}

\begin{figure}
	\includegraphics[width=0.5\columnwidth]{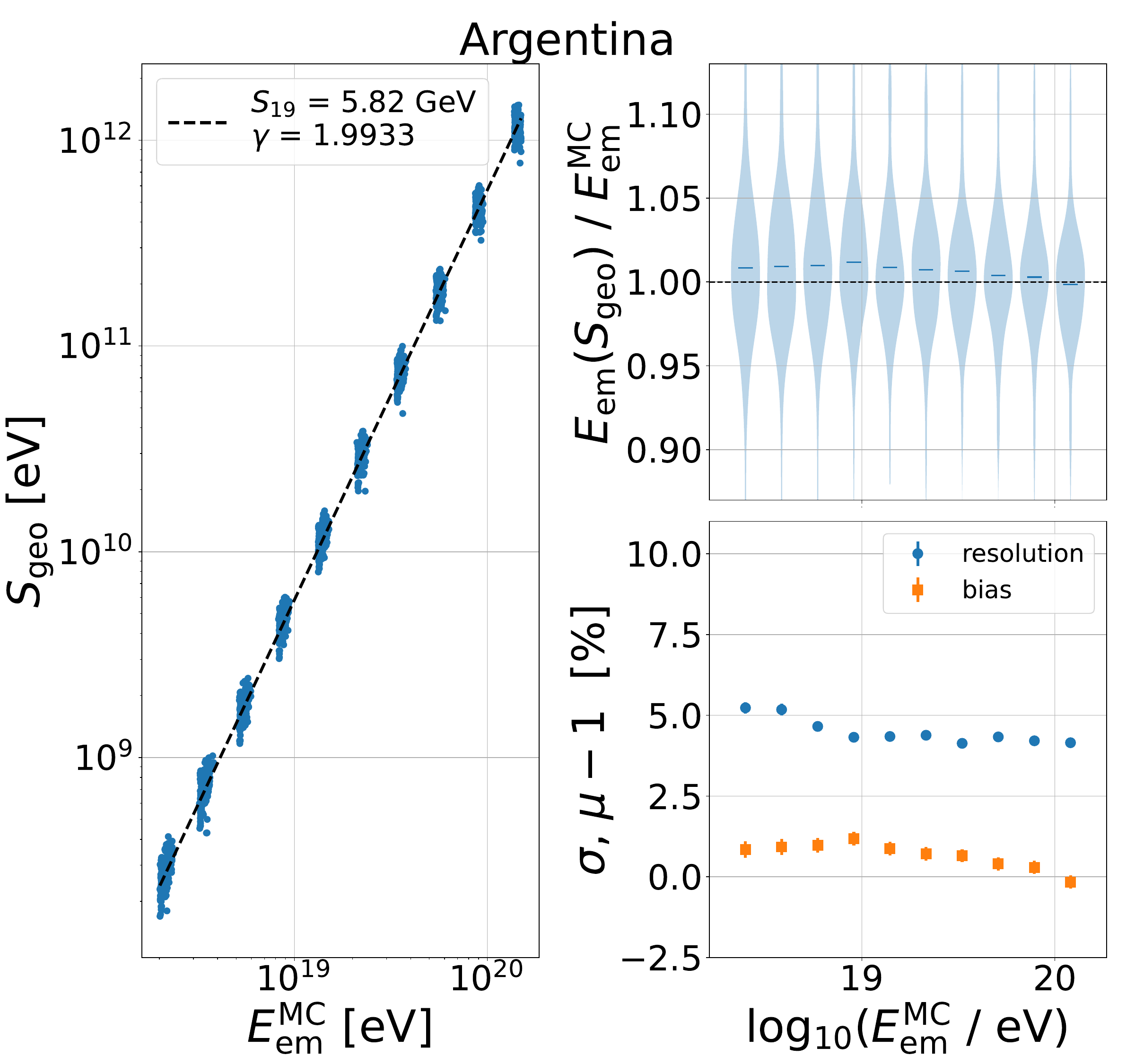}
     \includegraphics[width=0.5\columnwidth]{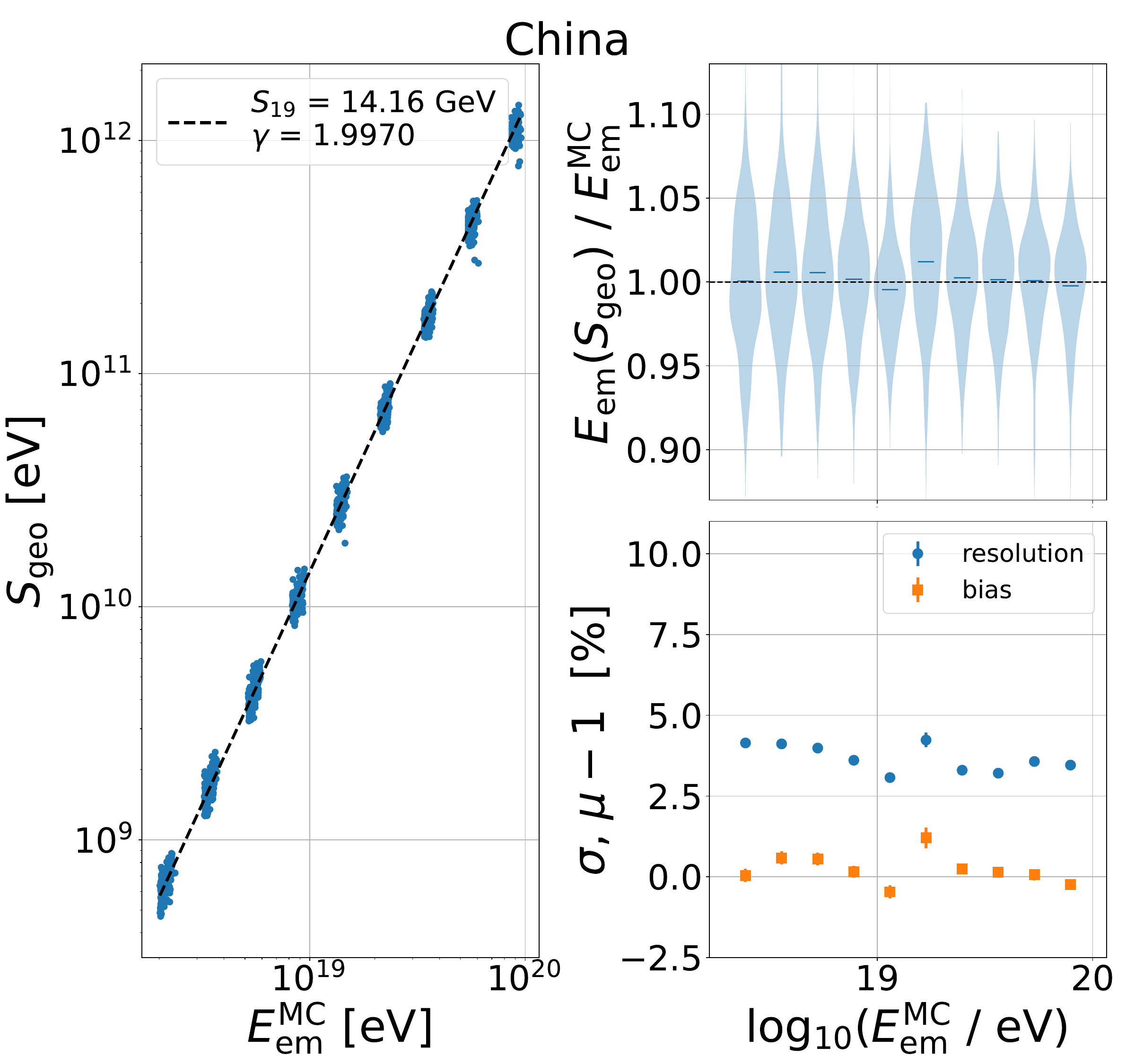}
    \caption{Reconstruction performances for the simulation sets without noise and star-shaped antenna layouts.
    The left column of each panel shows the correlation of the reconstructed $S_\text{geo}$ against the true $E_\text{em}^\text{MC}$.
    The dashed line represents the power law from Eq.~(\ref{eq:em}).
    The right columns of each panel display the deviation of the reconstructed $E_\text{em}$ from the Monte-Carlo truth.
    On top, we show the results as violin plots against $\text{log}_{10}(E_\text{em}^\text{MC} / \mathrm{eV})$.
    On the bottom, the corresponding resolution and bias of each bin is displayed.
    \textbf{Left:} Argentina.
    \textbf{Right:} China.}
    \label{fig:intrinsic}
\end{figure}

The first benchmark for our method is the intrinsic reconstruction performance on the simulation libraries with star-shaped antenna layout with which we tuned the parameters, with no added noise. 
Both libraries contain $\sim 4,000$ air-shower simulations with 240 antennas each, have zenith angles ranging from 65 to 85 degrees with a uniform distribution, uniformly distributed azimuth angles and primary energies binned log-uniform between $10^{18}$ and $10^{20}\,\mathrm{eV}$.
For this, the left panels of Fig.~\ref{fig:intrinsic} display the power law from Eq.~(\ref{eq:em}).
The right panels show for each site energy resolutions $<5\%$ with negligible biases.

Further, we look at two different simulation sets of sparse antenna grids for the site in China.
The first set simulates GP300 with 300 antennas with $1\,\mathrm{km}$ spacing and an in-fill with $500\,\mathrm{m}$ spacing.
The second set simulates an array with 10,000 antennas with spacing of $1\,\mathrm{km}$ and no in-fill.
Due to the detection area of $10,000\,\mathrm{km^2}$, we  effectively neglect boundary effects, which should, however, play a minimum role for such a large detector.
For both sets, primary energy, zenith and azimuth angles are generated randomly.
The primary energies are log-uniform in $[10^{17}, 10^{20}]\,\mathrm{eV}$.
The zenith angles are uniform in $[65, 85]$ degrees and azimuth angles are isotropic.
We add simulated Gaussian noise at the level of $64\,\mathrm{\mu V/m}$ to the electric field traces to match the expected noise level in measurements.
In addition, we smear the electric field amplitudes by $7.5\%$, and apply a $5\,\mathrm{ns}$ time jitter to the traces.

\begin{figure}
	\includegraphics[width=0.5\columnwidth]{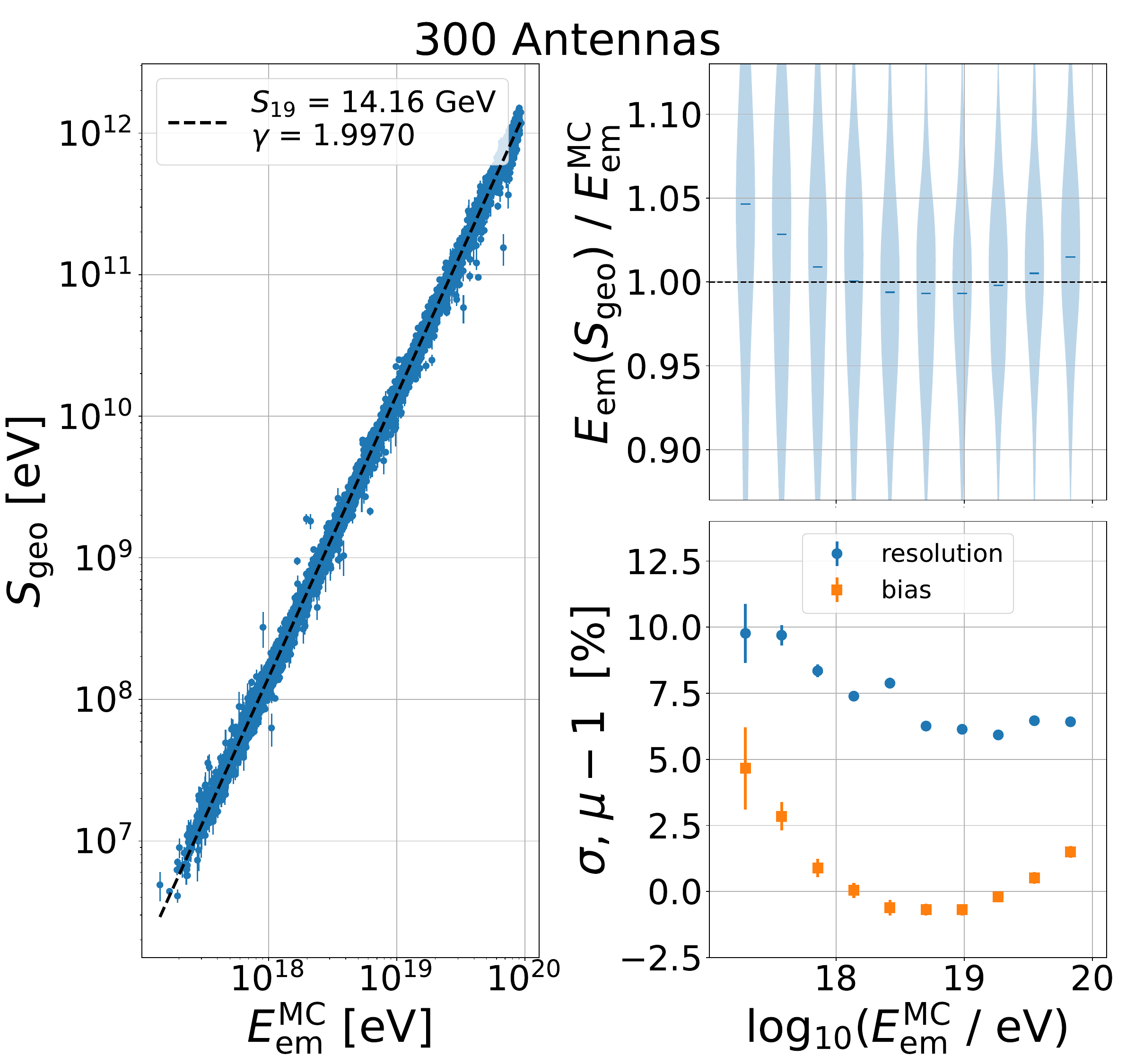}
     \includegraphics[width=0.5\columnwidth]{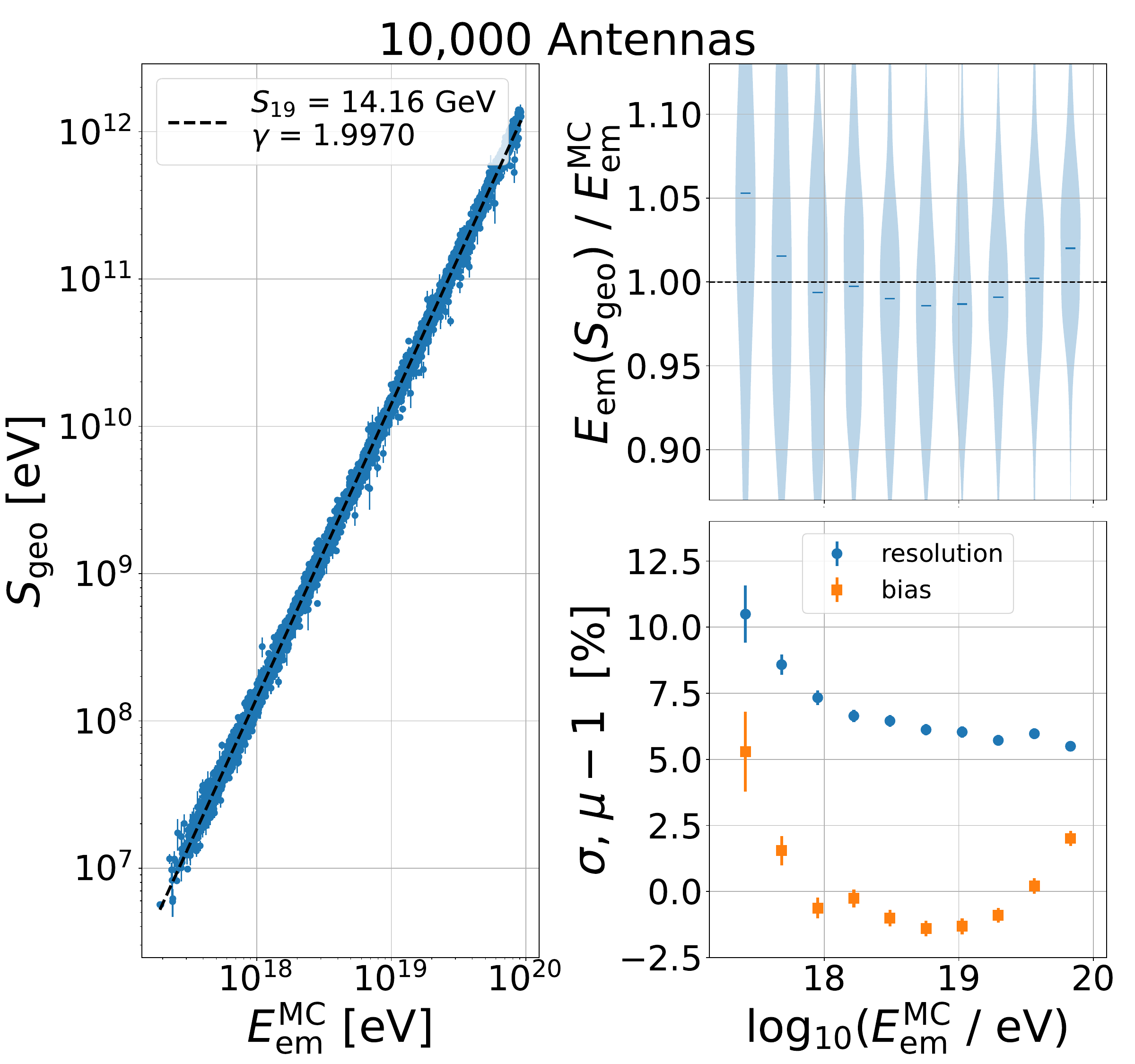}
    \caption{Reconstruction performances for the simulation sets with artificial noise and sparse antenna layouts.
    Both panels have the same properties as the panels in Fig.~\ref{fig:intrinsic}.
    \textbf{Left:} $15,000$ simulations for GP300 array.
    \textbf{Right:} $6,000$ simulations for $10,000\,\mathrm{km^2}$ array.}
    \label{fig:L1}
\end{figure}

Fig.~\ref{fig:L1} shows the reconstruction performance for the GP300 array (left), and for the $10,000\,\mathrm{km^2}$ array (right).
For both, the energy resolution is $<10\%$ with still negligible biases.
For low energies, where the signal amplitude approaches the noise level, the performance worsens slightly.

\section{Conclusions}
\label{sec:conclusions}

In this work, we have adapted the signal model from reference \cite{Model-paper} to the $50-200\,\mathrm{MHz}$ frequency band for both the site of Pierre Auger in Argentina and the GRANDProto300 site in China.
We have refitted the parametrisation of the charge excess fraction.
We have modified the geomagnetic lateral distribution function to account for the emission properties in the wider and higher frequency band.
For the low magnetic field strength in Argentina, these changes are sufficient to adapt the model.
For China, the almost 3x as strong magnetic field leads to a loss of signal coherence in the radio emission.
To address this, we have adapted the density correction used to calculate the corrected radiation energy $S_\text{geo}$.
For both sites, we find intrinsic energy resolutions $<5\%$ with negligible biases.
This indicates that it is possible to adapt the model to any magnetic field configuration.

In addition, we have applied our reconstruction to simulations of realistic, sparse antenna arrays, and with added artificial noise.
For both the 300 as well as the 10,000 antenna array, we achieve an energy resolution consistently $<10\%$ with no significant bias except at low CR energies where the radio signal strength approaches the noise level.


\let\oldbibliography\thebibliography
\renewcommand{\thebibliography}[1]{%
  \oldbibliography{#1}%
  \setlength{\itemsep}{1pt}%
}
{\footnotesize
\bibliographystyle{JHEPnotitle}
\bibliography{bibliography}
}

\section*{Acknowledgments}
This work is part of the NUTRIG project, supported by the Agence Nationale de la Recherche (ANR-21-CE31-0025; France) and the Deutsche Forschungsgemeinschaft (DFG; Projektnummer 490843803; Germany).
The authors gratefully acknowledge the computing time provided on the high-performance computer HoreKa by the National High-Performance Computing Center at KIT (NHR@KIT).
This center is jointly supported by the Federal Ministry of Education and Research and the Ministry of Science, Research and the Arts of Baden-Württemberg, as part of the National High-Performance Computing (NHR) joint funding program. HoreKa is partly funded by the German Research Foundation.

\end{document}